\documentclass{article}

\usepackage{arxiv}

\usepackage[utf8]{inputenc} 
\usepackage[T1]{fontenc}    
\usepackage{hyperref}       
\usepackage{url}            
\usepackage{booktabs}       
\usepackage{amsfonts}       
\usepackage{nicefrac}       
\usepackage{microtype}      
\usepackage{lipsum}
\usepackage{graphicx}
\usepackage{amsmath}
\graphicspath{ {./images/} }
\usepackage{apacite}

\title{Quantification of Pollen Viability in \textit{Lantana camara} By Digital Holographic Microscopy}

\author{
Vipin Kumar$^*$\\
National Institute of Plant Genome\\
Research, New Delhi 110067 India\\
\texttt{environment.vip14@gmail.com} \\
   \And
Nishant Goyal$^*$\\
Department of Physics\\
Indian Institute of Technology Delhi\\
New Delhi 110016 India\\
\texttt{phz218060@iitd.ac.in} \\
  \And
Abhishek Prasad\\
National Institute of Plant Genome \\ Research, New Delhi 110067 India\\
\texttt{abhishek1992@nipgr.ac.in}
  \And
Suresh Babu\\
School of Human Ecology,\\
Dr. B.R Ambedkar University Delhi,\\
New Delhi 110006 India\\
\texttt{suresh@aud.ac.in}
  \And
Kedar Khare\\
Optics and Photonics Centre\\
Indian Institute of Technology Delhi\\
New Delhi 110016 India\\
\texttt{kedark@physics.iitd.ac.in}     \And
Gitanjali Yadav\\
National Institute of Plant Genome\\
Research, New Delhi 110067 India\\
\texttt{gy@nipgr.ac.in}
}

\begin{document}
\maketitle
\def\thefootnote{*}\footnotetext{These authors contributed equally to this work.}\def\thefootnote{\arabic{footnote}}
\begin{abstract}
\textbf{Pollen grains represent the  male gametes of seed plants and their viability is critical for efficient sexual reproduction in the plant life cycle. Pollen analysis is used in diverse research thematics to address a range of botanical, ecological and geological questions. More recently it has been recognised that pollen may also be a vector for transgene escape from genetically modified crops, and the importance of pollen viability in invasion biology has also been emphasized. In this work, we analyse and report an efficient visual method for assessing the viability of pollen using digital holographic microscopy (DHM). We test this method on pollen grains of the invasive \textit{Lantana camara}, a well known plant invader known to most of the tropical world. We image pollen grains and show that the quantitative phase information provided by the DHM technique can be readily related to the chromatin content of the individual cells and thereby to pollen viability. Our results offer a new technique for pollen viability assessment that does not require staining, and can be applied to a number of emerging areas in plant science.}
\end{abstract}


\section{Introduction}
Pollen grains commonly appear as fine dust; each grain represents a minute body of varying form and size, produced in specialized male floral organs called stamens in seed bearing plants, and transported to the female structures for fertilization. Viability refers to the capacity of pollen grains to mature and then fertilize, followed by the ability to develop into seed and fruit \cite{shapiro1965pollen, faegri1989textbook}. Pollen analysis has long been known as a rigorous scientific method encompassing a diverse number of research disciplines including botany, paleoentology, geology, ecology, climatology, archaeology and many more \cite{faegri1989textbook, dafni2000pollen}. More recently, in the modern era, pollen have been recognised as vectors for transgene escape from genetically modified crops \cite{firmage2000field}. Furthermore, these tiny male reproductive units of seed plants have also been recognised for their importance in biological invasions \cite{bufford2014sterility}. Ecosystem threats by invasive plant species often pose a huge challenge, not only for conservation but also for the nursery, horticultural and landscape industries, leading to the necessity of exploring sterile non-invasive cultivars to replace fertile invasive ones \cite{burns2019plant}. In the following sections, we bring together literature that strings together pollen viability estimations, biological invasions and the importance of imaging techniques in pollen analysis. Following this discussion of prior literature, we describe a novel quantitative phase imaging methodology for pollen using a digital holographic microscope system (DHM). DHM is an emerging modality where the natural refractive index contrast of the sample is imaged in a quantitative manner as explained in detail later. The DHM system we use works with a dual mode operation, so that, a single pollen can be imaged in the usual bright-field mode as well as in the quantitative phase mode by simple switching of illumination without disturbing the sample. This allows us to establish phase map characteristics of viable and non-viable pollen in a quantitative manner. A DHM system was used for pollen refractive index tomography using multiple hologram recordings of a rotating sample followed by Radon transform inversion \cite{charriere2006cell}. A combined fluorescence and digital holographic microscopy system was demonstrated with pollen grains as test sample \cite{shaffer2012single}. Pollen imaging in quantitative phase imaging mode has also received attention in environmental sciences literature \cite{van2004method, berg2011digital, wu2018lensless, sauvageat2020real, kemppinen2020imaging}.
In this work a novel single-shot full resolution DHM system that uses a sparse optimization based phase recovery \cite{khare2013single, singh2017single, singh2018single, mangal2019unsupervised} has been employed to image a large number (more than 500) of individual pollen grain from \textit{lantana camara}. The full resolution (with 40x, 0.75 NA) image plane holography of the pollen as performed here is clearly seen to bring out morphological features of viable and non-viable pollen in their hydrated state. We further provide initial evidence that quantitative phase mode of imaging may enable pollen imaging and classification without the need of conventional staining procedures.  

The paper is organized as follows. In Section 2, we describe the importance of pollen viability studies to plant sciences and briefly review the existing techniques for viability estimation. Section 3 details our sample preparation and imaging methodology with some additional details on the single-shot phase reconstruction algorithm. An illustration showing visual morphological differences in different types of pollen as evident from their quantitative phase maps is presented. This is followed by Section 4 with quantitative measurements and statistics performed on the phase images of over 500 \textit{lantana} pollen grains. An illustration of phase imaging for unstained pollen grains is also presented.
Finally in Section 5 we provide concluding remarks and our scope for future work including new directions being undertaken in our laboratory.

\section{Pollen viability and its importance}
Viable pollen is important for species dispersal, fitness, and survival of the next plant generation. It is also essential for directed plant breeding and, consequently, crop improvement. Pollen viability comprises different aspects of pollen performance such as fertilization ability, germinability, and stainability\cite{dafni2000pollen}.  Pollen viability can be affected by drought/dehydration, heat stress and UV-B radiation. These factors can play a role after pollen dehiscence, when pollen is exposed to the environment, or even before, during pollen development inside the anther. Viability may also be species-specific and dependent on pollen physiology, or presence of specific structural modifications. A complicating factor in these assessments is the lack of standardized protocols and experimental conditions for viability assessments as described below \cite{dafni2000pollen, firmage2000field}. 
\subsection{Techniques for Viability Estimation}
AS pollination is the primary function of the pollen grain in a plant life cycle, one way to test
pollen viability is to use the pollen for pollination and subsequently analyze seed
set. However, this is time consuming and often not feasible, thus other methods are
frequently used to elucidate pollen viability, such as staining techniques, in vitro or in-vivo germination, as well as semi-in situ germination on the excised stigma (stigmatic germination), or impedance flow (IF) cytometry, but again, most of these are difficult to scale up and further confounded by incompatibilities, post-fertilization barriers and limited measurability, factors that restrict the accuracy of these tests \cite{dionne1958staining, dafni2000pollen}. 

Vital staining is by far the fastest and most commonly used method of pollen viability estimation and this involves visualization of specific compounds, contents, or cellular compartments. For example, Potassium iodide, aniline blue, and acetocarmine stains bind to starch, callose, and chromatin, respectively, and the absence of colors indicate non-viable pollen. \cite{stanley1974viability}. The Alexander stain discriminates aborted pollen grains from non-aborted pollen grains by staining the cytoplasm in red, and the cell walls green. In the absence of cytoplasm (non-viable grains), the green cell walls become visible, indicating lack of viability \cite{alexander1969differential}. The last major advance in this area was almost half a century back when the Heslop-Harrisons developed a viability test based on fluorochromatic reaction based membrane integrity and enzyme activity \cite{heslop1970evaluation, heslop1977pollen, shivanna1981membrane}. Although staining methods offer the possibility to distinguish aborted and non-aborted fresh pollen, they often fail to discriminate different viability levels \cite{ge2011pollen}.

Refined viability estimations have been attempted by comparing and combining some of these techniques. For example, high correlations were found between viability results from IF cytometry and FDA staining in case of mature cucumber, sweet pepper, and tomato pollen\cite{heidmann2016impedance}. However, pollen viability assessment purely through microscopy has not been investigated up till now. Therefore, the present study aims to assess whether DHM can be used as a standalone method for checking viability of pollen. The term “pollen viability” has been used as an umbrella term describing the capacity of pollen to live, grow, germinate, or develop \cite{dafni2000pollen}. By manually comparing 500 individual pollen grains of \textit{Lantana camara}, we selected distinguishing features amongst over 30 parameters identified by the DHM technique. The results were compared with the pollen viability assessed by the fastest method (acetocarmine staining) and it was found that DHM can very successfully discern one from the other. 

\subsection{Invasive species pollen analysis; Case Study \textit{Lantana camara}}

Our interest in pollen analysis is an extension of our ongoing efforts to understand Invasive alien plants species (IAPS), particularly \textit{Lantana camara} \cite{mishra2021decision, chauhan2022ecological, davis2000eight}. IAPS are considered to be one of the major drivers of biodiversity loss, posing severe threats to ecosystem services, environmental quality and human health globally \cite{jones2018health, bartz2019assessing, pejchar2009invasive}. \textit{Lantana camara} (Wild Sage) is a small broadleaf flowering shrub within the Verbenaceae family, native to American tropics \cite{sharma1988review}. It is an extremely adaptable weed, found across a wide variety of ecosystems. Once Lantana has been introduced into a habitat, it spreads rapidly, and has already done so from it's native Central and South America to over 50 countries, making it one of the world's top ten invasives \cite{ghisalberti2000lantana}. 

Pollen viability data becomes important in context of both, species invasivity as well as habitat invasibility, because it can provide important information on successful establishment of a given species in a habitat, thereby serving as a measure of invasiveness \cite{jiang2022reproductive}. For this reason, production of seedless or sterile invasive plants would be beneficial both to the nursery/horticultural industry as well as the environment \cite{beck2012optimisation}. One way of identifying sterile plants is to estimate viability across populations but as mentioned in earlier sections, this is often time consuming on account of technique limitation. A study on 32 L. camara cultivars and breeding lines revealed that stainability, the main parameter used to determine viability, is influenced by ploidy levels, indicating a strong potential to develop genetically sterile cultivars \cite{czarnecki2014ploidy}. Other studies have also established that assessment of morphological and cytological differences among lantana varieties can help in measuring invasive potential and suitability for commercial production and landscape use \cite{steppe2019morphological}. To date, the only method faster than vital staining appears to be in pollen imaging, but this has not been addressed nor explored fully as described in the next section. 
\subsection{Imaging of pollen grains}

Traditionally, palynologists have used compound light microscopy (LM) for pollen identification and interpretation and scanning electron microscopy (SEM) for morphological comparisons and taxonomy \cite{jones2007comparison}. Doubtlessly, SEM offers far greater resolution, and has led to creation of new terminology for describing pollen ornamentation, numerical approaches to pollen sculpturing and exine architectures\cite{skvarla1989analysis}. However, sample preparation and the time needed to count, analyze, photograph and print the micrographs, and the consequent lack of scalability are the limiting factors in these approaches.

With the advancement of digital microscopy, palynology studies are becoming even less time-consuming and can generate more reliable data for species taxonomy, apart from saving hours spent on manual counts of pollen grains following the process of
staining to differentiate between viable and inviable pollen \cite{mishra2015palynological}). 
At the turn of the millenium, alternative digital
methods for counting pollen were devised, but often without regard to viable and invisible grains, and these protocols also had drawbacks of software specific to branded instruments. \cite{bechar1997image,aronne2001classification}. 

More recently, particle counters and binary 2D projection including shape-based Fourier descriptors and topological features, have been used for detecting and classifying real and artificial pollen grains, but the success varies by species, and is applicable only to species where size difference settings for viable
and inviable grains is larger than the natural variation in viable pollen size for that species \cite{kelly2002method, mudd2012image, akcam2022pollen}

In summary, the advantages of microscopy and advanced pollen imaging techniques were able to overcome limitations related to manual counting under a microscope using a haemocytometer as well as slides
with grids that accommodated a specified volume of the solution, all of which were extremely time consuming and labor intensive, requiring each sample to be counted twice (once for viable and
again for inviable pollen). Flow cytometry methods were then described to determine pollen viability based on
the nuclear DNA content in mature pollen grains but this is a fairly lengthy and expensive process not suitable
for a quick estimate \cite{bino1990flow}.
The particle counter overcame these limitations but required generation of preliminary size distributions for each species by staining and visual
counting, since it is necessary to have little or no overlap in the size distributions of inviable and viable grains.
Digital image processing appears to overcome size based differences and species limitations but has not been explored sufficiently widely, emphasizing the importance of another simpler method for pollen gain imaging in natural/hydrated state for assessing viability, as described in this work by means of the DHM.
 
\begin{figure}
    \centering
    \includegraphics[width=0.95\textwidth]{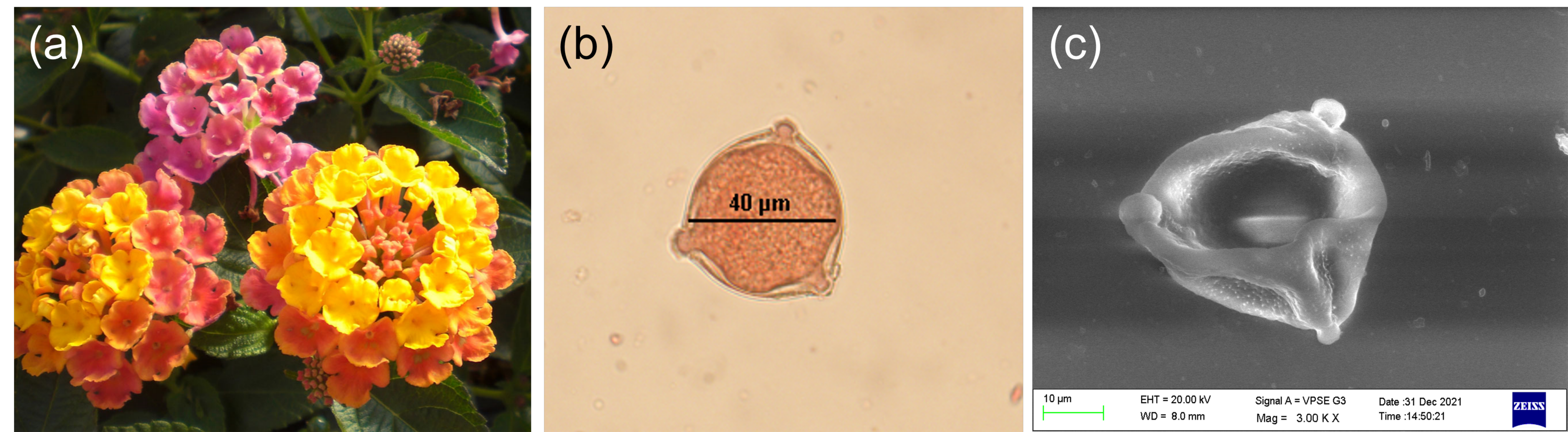}
    \caption{(a) Lantana camara flowers, and typical lantana pollen grain images observed with (b) an optical microscope, and (c) a 20kV scanning electron microscope. Note the surface texture and tricolpate nature of the pollen grain}
    \label{fig:lantana}
\end{figure}

\section {Methodology}
\subsection{Sample Collection and Pollen Analysis}
\subsubsection{Sample Collection and Staining}
\textit {Lantana camara} flowers were collected from  Hauz Khaz Rose Garden (28°32'52.09"N/ 77°11'27.77"E) which is a part of Delhi Ridge forest. The sample collection was done from January-March 2022. Fresh flowers were plucked early morning, which were then directly transferred to the laboratory for analysis. Pollens were extracted from florets and transferred on slide using tapping and squashing method as described here. For stain preparation, 2g of Carmin powder was dissolved in 95ml of glacial acetic then added distilled water to make 100ml solution. The solution was boiled, filtered and stored in room temperature. Flower clusters (inflorescences) were collected when the clusters each had at least one flower partially open. Pre-dehiscent anthers were removed from unopened flowers and placed in a 1.5-mL Eppendorf tube with  $\approx$ 100 $\mu$L of acetocarmine stain. Anthers were stained and rinsed three times with distilled water. Care was taken not to burst anthers while rinsing them. Rinsed anthers were squashed in 50 $\mu$L of glycerol (diluted with 20 percent water) on a glass slide. Stained anthers were transferred onto a microscope slide and covered with a coverslip. The slide was gently tapped and pressed to release pollen grains out of anthers. Pollen viability was observed and recorded manually through a microscope after an interval period of 5-10 minutes, as described in the next section.
\subsubsection{Microscopy based Pollen Viability}
\textit {Lantana camara} pollen grains  were observed a few minutes after staining as described above, under a bright field Nikon Eclipse E200 upright microscope, using 40× magnification objective for initial preparation. Photos of pollen grains were taken using an IDS $\mu$Eye 3070CP camera mounted to the microscope. Images of pollen grains were later examined on a MacPro OSX Workstation. Pollen grains stained to dark red were considered stainable. As shown in Figure \ref{fig:lantana}(b), uniformly round, non-wrinkled, brightly fluorescing pollen grains were considered viable, while non-fluorescing or lightly fluorescing and wrinkled or deformed pollen grains were considered non-viable. Manual scanning of over 1200 pollen grains in this manner and removal of clumped grains led to the selection of 500 clearly defined viable and nonviable pollen grains. 

\subsection{Preparation of pollen grains Scanning Electron Microscopy (SEM)}
The surface texture of pollen grains was examined using a SEM microscope and this was used to validate species identity. Specimens of viable \textit {Lantana camara} pollen grains were prepared for SEM by mounting individual undehydrated pollen grains from ddH$_2$O onto a double-sided adhesive carbon disk that was attached to an SEM stub. The stub was then coated with gold-palladium using a sputter coater, and specimens were viewed using a computer driven Leica TCS-SP2 Confocal Laser Scanning Microscope (AOBS TCS-SP2) at 15kV and 20 kV resolution. The pollen unit was found to be monads or tetrads, with psilate type sculpture and colpi largely tricolpate as shown in Figure \ref{fig:lantana}(c). Quantitative parameters such as polar diameter, mesocolpium distance, equatorial dimensions, aperture size, spine diameter, and exine thickness were calculated using the open-source ImageJ software (https://imagej.nih.gov/ij/). This was followed by DHM imaging of all 500 pollen grains, as described in the next section.

\subsection{Digital holographic microscopy and its application to pollen imaging}
\begin{figure}[htbp]
    \centering
    \includegraphics[width=\textwidth]{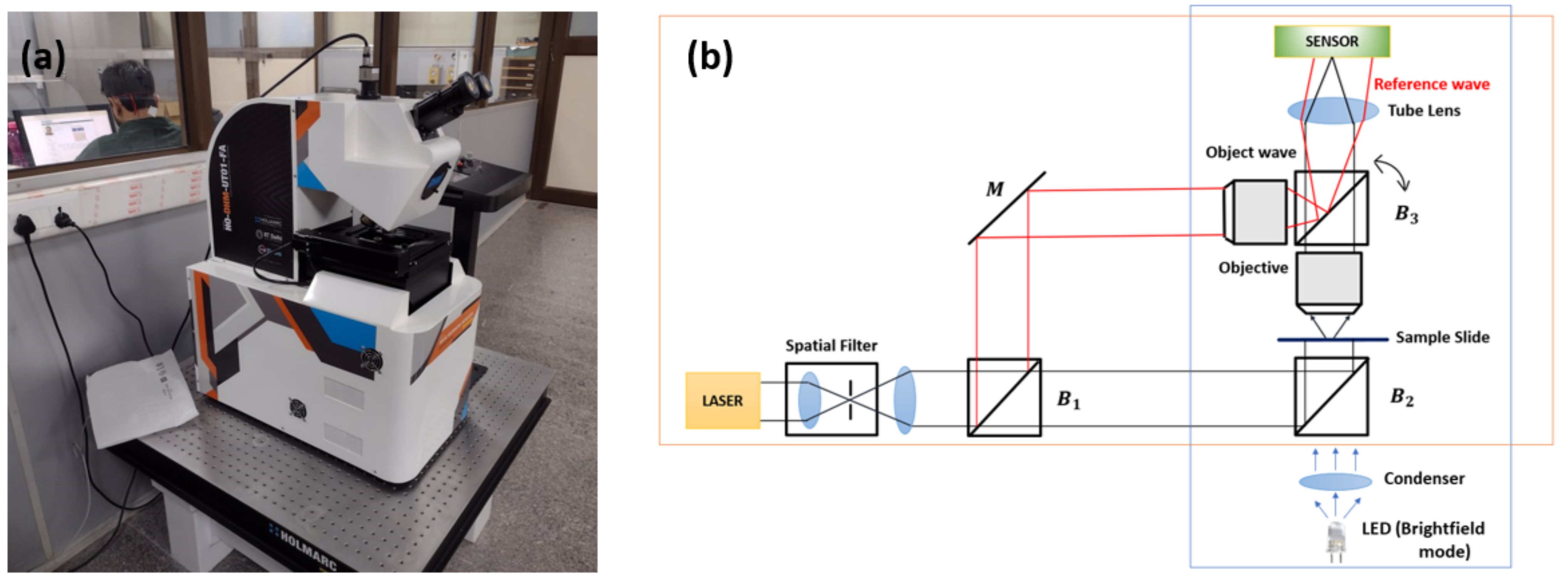}
    \caption{Digital holographic microscope system used in the present study, (a) System photograph, (b) Nominal optical layout of the system. $B_1, B_2, B_3$: Beam-splitters, $M$: plane mirror. The beam-splitter $B_3$ can be rotated with an adjustable screw in order to introduce tilt in the reference wave as required for off-axis hologram recording. The system is fitted with a laser source for phase imaging and LED illumination with condenser for bright-field illumination. By switching between the two illuminations one can record an image plane digital hologram or a bright-field image of the sample in the same position.}
    \label{fig:DHM}
\end{figure}
Digital Holographic Microscopy (DHM) is an interferometric imaging modality that uses a low power (few mW) coherent laser illumination. The DHM system shown in Fig. \ref{fig:DHM} that was used in the present study consists of a Mach-Zehnder type interferometer, where a coherent laser beam is first split into two parts using a beam-splitter. 
One of the beams is transmitted through the pollen sample and forms the image field $O(x,y)$ at the detector as formed by an infinity corrected imaging system. The second beam takes an identical path without any sample and forms a tilted plane reference beam $R(x,y)$ at the detector plane. Since the two optical fields $O(x,y)$ and $R(x,y)$ have originated from the same coherent laser beam and travel near identical path lengths up to the detector, they are temporally coherent and produce interference fringes with good contrast at the detector. A low power light source is sufficient for this modality as we are not exciting any fluorescence labels but simply transmitting the light beam through a semi or fully transparent sample. The interference pattern (or digital hologram) $H(x,y)$ recorded on the pixelated array detector may be mathematically described by:
\begin{align}\label{eq:hologram}
    H(x,y) &= \lvert R(x,y) + O(x,y) \rvert^2 \nonumber \\&= \lvert R \rvert^2 + \lvert O \rvert^2 + R^*O + RO^*.
\end{align}
The * operation above is the complex conjugation of the wave functions associated with the object and reference waves. In the present case, the image plane type holography is implemented where $O(x,y)$ represents the image field of the pollen object recorded using a 40x infinity-corrected imaging system.  A laser source is embedded in the arrangement to sustain a good contrast in the fringe pattern. The sensor captures the Hologram pattern $H(x,y)$.

The image field $O(x,y)$ at the detector plane is a complex wave function and contains both amplitude and phase part. The amplitude part signifies the information regarding the attenuation or absorption of the illuminating light intensity when passed through the pollen sample. If the sample is stained, the space-varying amplitude information typically signifies the absorption of the illuminating beam by the sample. For unstained samples, there is minimal amplitude contrast. This is due to the fact that unstained samples act like transparent objects. Typical bright-field microscopes record only the $|O(x,y)|^2$ term to produce amplitude-contrast images. The important feature of a DHM is that it is able to record and numerically reconstruct the phase $\phi(x,y)$ of the image field as well. The phase map $\phi(x,y)$ may be interpreted as a projection:
\begin{equation}\label{eq:refind}
    \phi(x,y) = \frac{2\pi}{\lambda} \int  n(x,y,z)\ dz
\end{equation}
Here $n(x,y,z)$ represents the relative refractive index in the 3D space of the pollen sample relative to the surrounding medium. In other words, the phase function $\phi(x,y)$ represents the optical path length for the illuminating laser beam through the sample pollen. Note that the refractive index $n(x,y,z)$ is an inherent material property of the pollen grain being imaged. This material property is thus recoverable from a DHM system but not accessible with a bright-field system. Since the refractive index $n(x,y,z)$ has been integrated over the depth of the sample (or the propagation direction $\hat{z}$ of the light beam through the sample, the phase map also implicitly embeds the morphological information about the pollen grain. As an aside, we remark that the more commonly used phase contrast mode in microscopes does utilize the phase property of light to enhance the qualitative visual appearance of the samples. Unlike phase contrast mode, the DHM systems uniquely provide quantitative phase information which is much more meaningful for any classification task as in the present work.  

As may be understood from previous discussion, DHM is a computational microscope. The phase information of interest here is not directly measured on the array detector. Phase can however be estimated computationally using numerical processing of the recorded hologram pattern $H(x,y)$ at the detector plane. The speciality of our DHM system is that it provides single-shot phase imaging at full diffraction limited resolution. With current advances in the CMOS array sensor technology, a good contrast fringe pattern may be recorded in just a milli-second exposure time when used with a few mW laser source for illumination as in our case. The single-shot operation is advantageous as the system is not very sensitive to the surrounding vibrations during the exposure time.

Single-shot off-axis holograms as in our study are conventionally reconstructed using the Fourier transform-based method \cite{takeda1982fourier,kreis1986digital}. This method inherently involves a low-pass filtering operation and as a result the recovered object wave function $O(x,y)$ often has lower resolution compared to the diffraction-limit of the infinity corrected imaging system used. In image plane holography this loss of information results in blurring of edges in the image field that are critical to perception of the quality of the recovered phase image. Contrary to this, the bright-field images recorded using the same infinity corrected imaging system allow full diffraction-limited resolution. Hence the phase information recovered using Fourier transform method lacks resolution competency against the conventional bright-field images. Other methods such as phase-shifting methods \cite{creath1988v,yamaguchi2006phase}, can deliver the expected full resolution, but require multiple hologram recording with much stringent vibration isolation making the DHM system costs high. In order to retrieve the phase information with full pixel-resolution from a single-shot off-axis hologram, we use a sparse optimization method that we developed over last several years of effort \cite{khare2013single}. The resolution and phase accuracy advantages of this optimization method are already well established in prior literature.

In brief, in the sparse optimization methodology, the reconstruction of the complex object wave $O(x,y)$ is formulated as an optimization problem where we minimize a cost function given as:
\begin{equation}\label{eq:cf}
\begin{aligned}
    C(O,O^*) & = C_1 + C_2\\
             & = \lvert\lvert H - (\lvert R \rvert^2 + \lvert O \rvert^2 + R^*O + RO^*) \rvert\rvert^{2}_{2} + \alpha \psi(O,O^*)
\end{aligned}
\end{equation}
Here, the $\lvert\lvert...\rvert\rvert_2^2$ represents the squared L2-norm of the quantity inside. The first term in the cost function represents the least square data fitting between the measured hologram data $H(x,y)$ and the interference pattern formation model, while the second term $\psi(O,O^*)$ represents an image domain constraint. The constant $\alpha$ denotes a positive regularization constant that determines the relative weight between the two terms of the cost function. In the present work, we used the total variation (TV) penalty function as a constraint for the optimization problem. TV is a popular edge-preserving penalty function that has been widely used in the imaging literature which is suitable for the present image plane holography data. The total variation penalty is defined as:
\begin{equation}\label{eq:TV}
    \psi(O,O^*) = \iint \lvert \nabla O(x,y) \rvert \ dx \ dy
\end{equation} 
Further, we utilize an adaptive alternating minimization scheme \cite{sidky2008image} for reconstruction of the complex valued object field $O(x,y)$ that does not involve empirical selection of any specific regularization parameter $\alpha$. The adaptive alternating minimization scheme is already described in prior publications \cite{singh2017single}. 

Since object data is complex-valued, Wirtinger derivatives with respect to $O^*$ are evaluated to obtain the steepest descent directions in which the solution progresses for any given iteration. The Wirtinger derivatives of two terms $C_1$ and $C_2$ in the cost function $C$ in Eq. (\ref{eq:cf}) are given as:
\begin{equation}\label{eq:delC1}
    \nabla_{O^*} C_1 = -2\  \Big[  H - \lvert R + O \rvert^2 \Big]\ .\ (R + O)
\end{equation}
and
\begin{equation}\label{eq:delC2}
    \nabla_{O^*} C_2 = -\ \nabla . \Big[ \frac{\nabla O}{\sqrt{\lvert \nabla O \rvert^2 + \epsilon^2}} \Big]
\end{equation}
Here $\epsilon$ is a small positive number ($\approx 10^{-10}$) used to prevent zero division. Since this optimization process operates completely in the image domain, a region of interest can be suitably chosen to allow a full-resolution recovery of $O(x,y)$ over a region of interest of the recorded image plane hologram $H(x,y)$. In other words, a user can easily select a region of interest near a pollen grain for the phase map recovery. In the present study, a $512 \times 512$ pixel area is used for phase reconstruction of individual pollen. 

Using the optimization procedure, the recovered complex object wave field delivers the spatial phase map $\phi(x,y)$ which itself is defined by the arctangent of the ratio of imaginary and real parts of $O(x,y)$, \begin{equation}\label{eq:phi}
    \phi(x,y) = \arctan \Bigg( \frac{Im\ [O(x.y)]}{Re\   [O(x,y)]} \Bigg)
\end{equation}
We emphasize that since the arctangent function is defined only for the range $[-\pi,\pi]$, the phase map $\phi(x,y)$ retrieved from the optimization procedure is wrapped in its form. This phase map must be unwrapped by some means to further relate with the sample characteristics accurately. In order to unwrap this phase map and derive the full physical information, a 2D unwrapping algorithm has been used that is based on the transport of intensity equation (TIE) method \cite{pandey2016two}.
The unwrapped phase function can be related to the refractive index of the pollen grain via the definition in Eq. (\ref{eq:refind}).

It is important to note that our DHM system (Make: Holmarc Opto-Mechatronics, Kochi, India, Model: HO-DHM-UT01-FA) works in dual mode, i.e., bright-field mode and holographic mode. This is depicted in Fig. \ref{fig:DHM}. The user can switch between the laser or LED illuminations for recording an image plane hologram or a bright-field imaging of the sample without physically disturbing it. Since the phase imaging is not a familiar modality among plant science researchers, the bright field mode can help locate the individual pollen (viable/non-viable) over the sample slide which may be focused appropriately. The holographic mode which essentially involves switching of the illumination can then record an image plane hologram of a pollen in the same focused position. Numerical reconstruction process applied to a recorded hologram then provides the user with the quantitative phase information $\phi(x,y)$ associated with each pollen grain.
Since the phase map represents optical path length, it is a common practice to render it as a surface plot to provide a sense of depth information regarding the object being imaged.
\section {Results}
\subsection{Quantitiative Phase Imaging of Lantana Pollen }
\begin{figure}[tbp]
    \centering
    \includegraphics[width=0.8\textwidth]{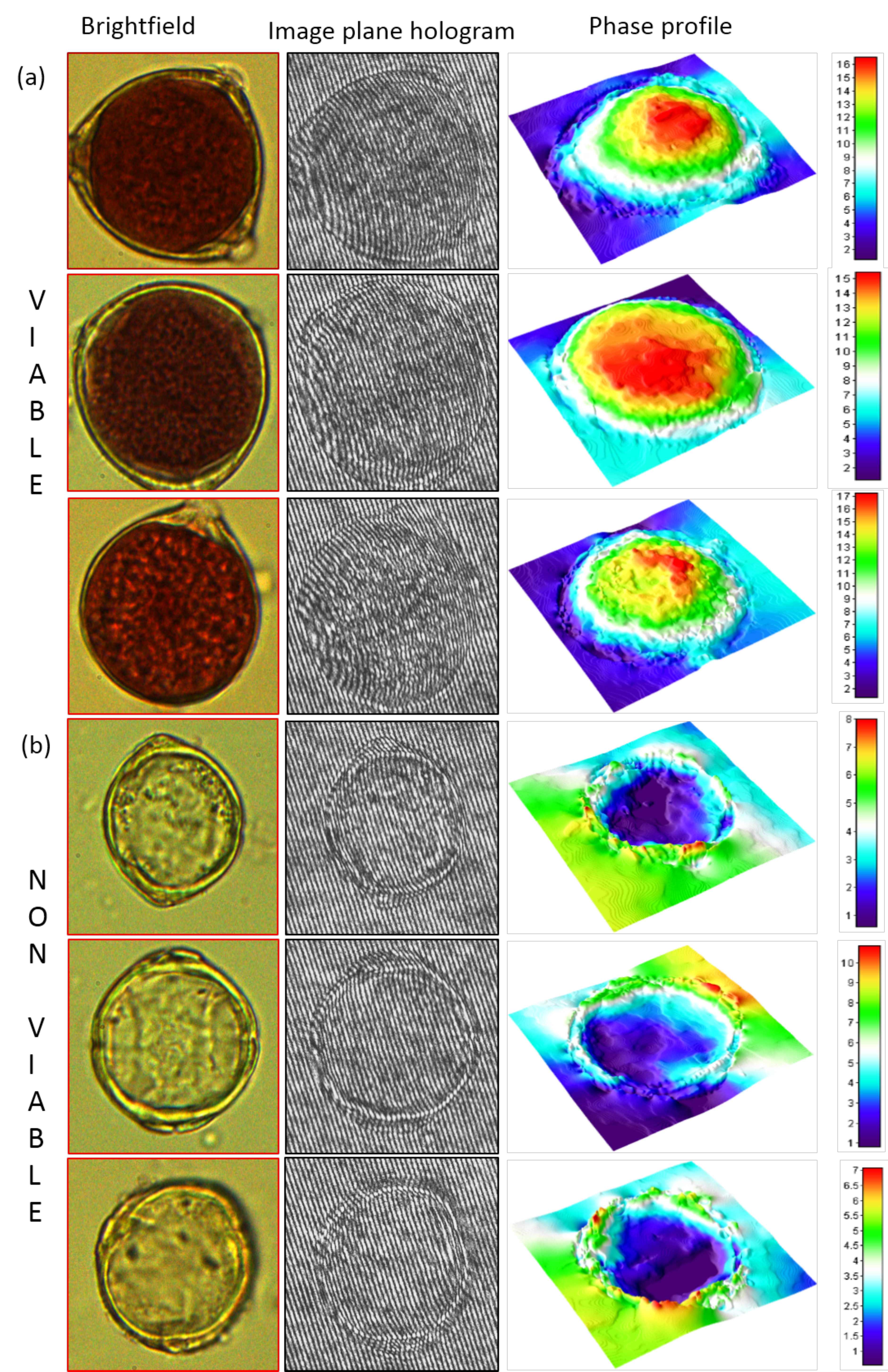}
    \caption{Brightfield Image, Recorded Hologram and 3D rendered phase map for (a) viable pollen, and (b) non-viable pollen. Note the ridges and fissures in the image plain hologram that are reflected as distinct layers in the phase profile}
    \label{fig:panel}
\end{figure}

For illustration, in Fig. \ref{fig:panel} we show the bright-field image, the recorded hologram (or interference pattern), and the 3D rendered phase map for a few select pollen of different types from $500$ individual lantana pollen images recorded in the bright-field and phase modes. We first make a few qualitative observations from the representative images. From the brightfield images, viable and non-viable pollen can be easily identified since the viable pollen show red color due to the identifying stain. The off-axis image plane hologram of the same pollen grain is depicted next which clearly shows bending of interference fringes at the location of the pollen grain. The corresponding phase map derived from the fringe pattern is depicted in the rightmost column as a surface plot as the phase map represents the optical height map of the pollen grain. It is interesting to observe that for pollen grains which show red color as in Fig. \ref{fig:panel}(a), the corresponding phase map plotted as a surface map nominally has a hemispherical shape. On the other hand the colorless pollen shown in Fig. \ref{fig:panel}(b), the phase map is nearly flat (as also seen from nearly straight line nature of the interference fringes within the pollen area). As per interpretation of the phase map as per Eq. (\ref{eq:refind}) and Fig. \ref{fig:panel}(a), (b), we observe that the main difference between the viable (red-colored) and non-viable (colorless) pollen is that the viable pollen contain extra material whose presence is detected in the form of the distinct hemisphere-like phase maps. The observation agrees with the prior knowledge that stain color represents sugars and DNA that absorb the dye inside the viable pollen. These observations suggest that the phase map as provided by a DHM system provides a successful and novel parameter for distinguishing viable or non-viable pollen. This methodology can be further established by performing numerical measurements on phase maps of a number of pollen and observing the corresponding statistics, as we demonstrate in the next section.

\begin{figure}[tbp]
    \centering
    \includegraphics[width=0.8\textwidth]{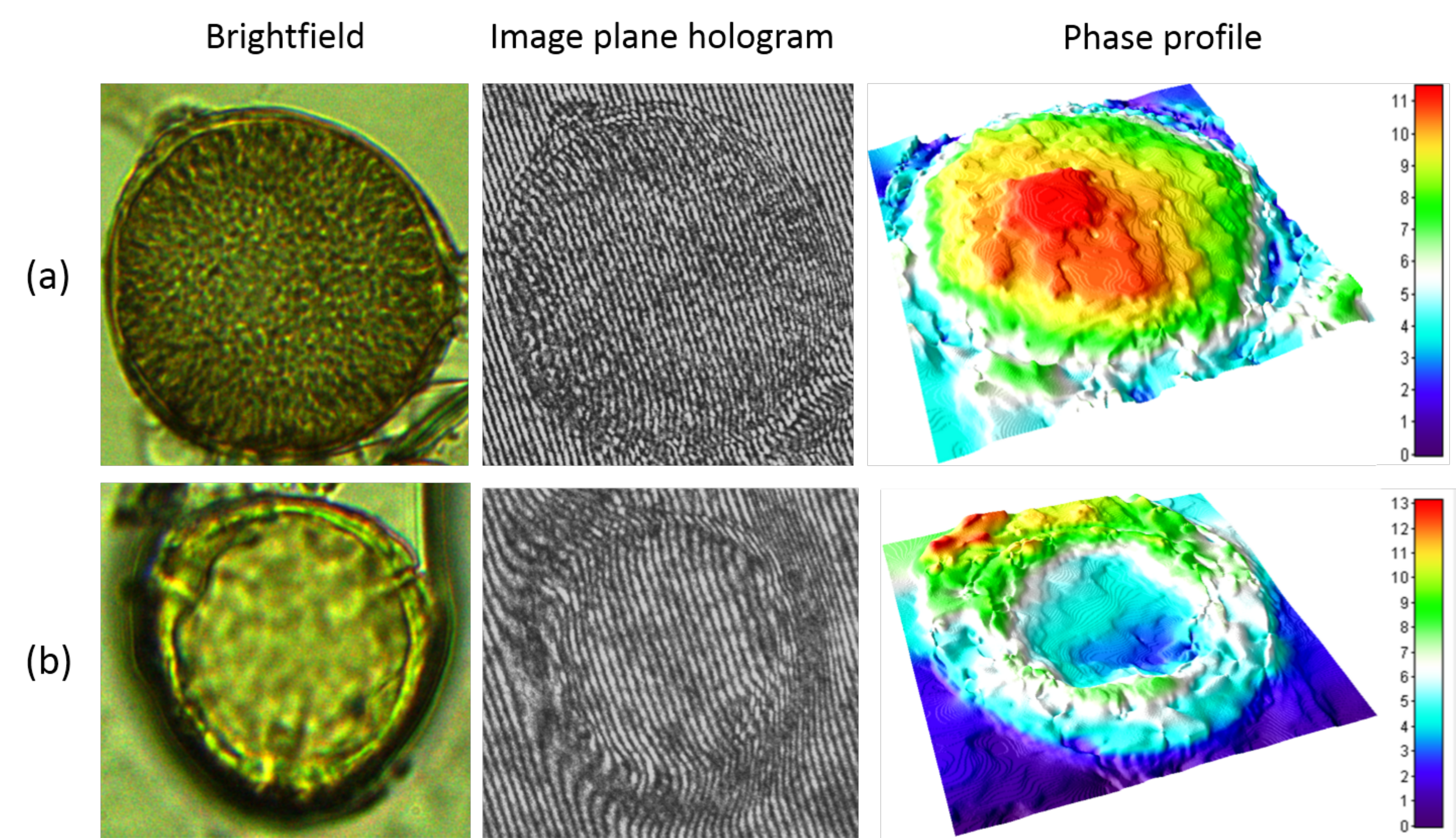}
    \caption{Illustrative Bright-field image, image plane hologram and 3D rendered phase map for two unstained pollen samples. The top and bottom rows represent viable and non-viable pollen with mean phase values 8.72 and 4.26 respectively.}
    \label{fig:unstained}
\end{figure}

\subsection{Quantitative measurements on bright-field and phase images}
\begin{figure}[htbp]
    \centering
    \includegraphics[scale=0.6]{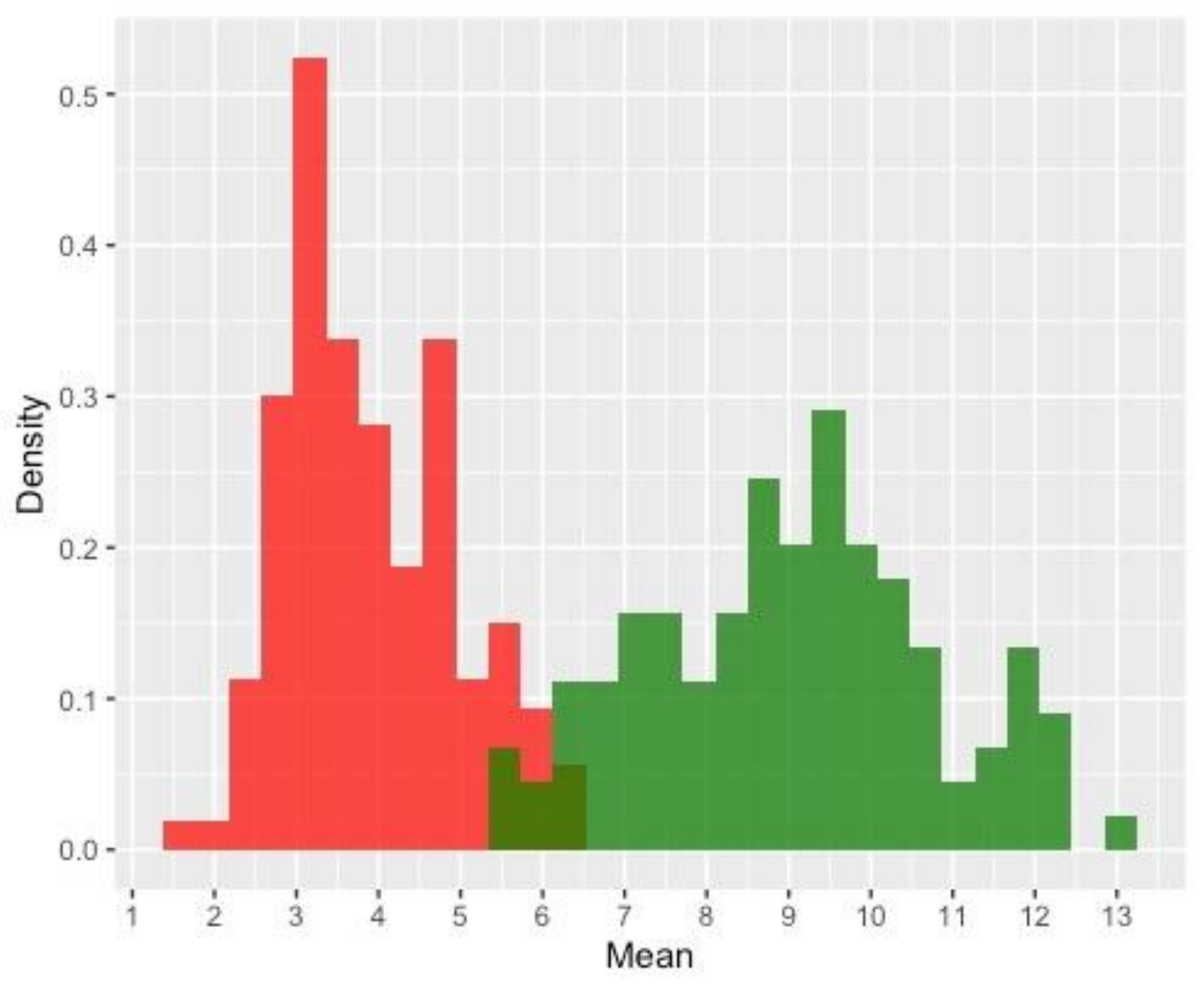}  
    \caption{Normalized histogram of mean phase values for two different classes of pollen. The red and green colors represent histograms for non-viable and viable pollen respectively.}
    \label{fig:histogram}
\end{figure}

The dual mode nature of our DHM system allows us to record image of a single pollen in both bright-field and quantitative phase modes without disturbing the sample. Since the two images are recorded with the same array sensor, it is very easy to identify the same pollen in images from the two modalities as is evident in Fig. \ref{fig:panel}. The images can be read in a standard open-source image processing platform like ImageJ. In our work we collected $500$ pollen images in both bright-field and quantitative phase modes. The bright-field image provides 2-dimensional parameters associated with the pollen such as area, perimeter, etc. The phase image on the other hand provides additional quantitative parameters such as mean height, optical volume, etc. In Fig. \ref{fig:histogram} we show probability density (or normalized histogram) of mean phase values of the $500$ pollen for which numerical measurements have been made. Mean phase value can be obtained by dividing the integrated phase values within the pollen by the area of the pollen grain. The histogram is labelled with 2 colors corresponding to the viable and non-viable pollen (as were clearly seen from presence or absence of the red stain in the bright-field images). The statistics of the viable and non-viable pollen is described in Table 1. The two sample t-test performed on the two classes of pollen has a p-value that is much lower than $0.05$ suggesting clearly different class means. 
\begin{table}[htbp]
 \caption{Statistics of mean phase values within pollen area for non-viable and viable pollen}
  \centering
  \begin{tabular}{llll}
    \toprule
     \textbf{Pollen type} & \textbf{Number of samples}    & \textbf{Mean phase} & \textbf{Standard deviation}             \\
    \midrule
    Non-viable   & $256$      & $3.90$ & $1.24$         \\
    Viable   & $252$     & $9.01$   & $2.17$      \\
    \bottomrule
  \end{tabular}
  \label{tab:table}
\end{table}
With the observation of clear distinction in phase maps for the viable and non-viable lantana pollen as described above, we further decided to examine the phase maps for unstained pollen. A fresh pollen sample was thus prepared without staining and the pollen were imaged with water as surrounding medium. Two illustrative pollen images for this unstained case are shown in Fig. \ref{fig:unstained} (a), (b).  It is not straightforward to distinguish these pollen in the bright-field mode in absence of the red stain. Their corresponding phase maps however suggest a possibility of distinguishing them. The mean phase value for pollen in Fig. \ref{fig:unstained}(a) is 8.72 and that for pollen in Fig. \ref{fig:unstained}(b) is 4.26. The numerical mean phase values are well withing the statistics and histograms presented in Table 1 and Fig. \ref{fig:histogram} respectively for the two pollen classes. This finding also suggests that the staining procedure does not significantly alter the refractive index of the cytoplasm material. A detailed study will be required in future to establish a label-free method based on quantitative phase imaging for pollen classification. 

Finally we remark that a simple visual inspection based method for pollen classification also seems to be possible without actual phase reconstruction. In particular we note from Fig. \ref{fig:panel} and Fig. \ref{fig:unstained} that viable pollen contain additional material in the pollen volume and as a result the interference fringes appear curved within the pollen boundary. On the other hand, for non-viable pollen which have lost the cytoplasm material, the interference fringes are nearly straight lines within the pollen area. Our main aim in this article is to establish suitability of the quantitative phase imaging modality for image based pollen classification. A method based on observation of nature of fringes can become an interesting future topic to be explored. We further mention that in the present study we mainly used the mean phase values within the pollen area as this parameter was sufficient to distinguish between the two pollen types. The phase images however clearly have much more textural details which may become useful in future detailed studies in this topic.   
\section{Conclusions and future directions}
Pollen viability refers to the capacity of pollen to get mature and then fertilize, followed by development into seed and fruit. Quantification of viability enables estimation of male reproductive potential of entire species, cultivars and populations. As described in the introduction, Several biotic and abiotic factors can affect pollen viability. This feature is often used in ecological, geological and genetic studies, and therefore estimation of viability among pollen samples of a particular species from a given region is often a critical prerequisite for a diverse array of research strategies in plant sciences. However, estimation of viability can be a time consuming and a low scale process, even the fastest methods available today require overnight dye based staining. 

In this work we describe a quantitative imaging technique using a single-shot full resolution digital holographic microscope (DHM) system which is able to image pollen in hydrated state. The most interesting aspect of this work is in our preliminary data that suggest the possibility of quantifying viability of pollen from unstained pollen samples, thus requiring minimal wet-lab processing for sample preparation. The DHM system also provided high resolution (full diffraction-limited) images of the same pollen grain in bright-field as well as quantitative phase modes, allowing us to establish a correspondence between pollen that show (or do not show) coloration due to vital staining and their corresponding quantitative phase maps. The natural refractive index of the sample is used as a contrast mechanism in the quantitative phase mode. As a result the quantitative phase maps of viable and non-viable pollen show a clear distinction based on presence or absence of the cytoplasmic material inside the pollen. The simplicity of our imaging methodology allowed us to image over 500 pollen grains making our findings statistically significant. 

In this paper, we focus on application of pollen viability to invasion biology by testing our method on pollen grains of the
invasive \textit{Lantana camara}, a well known plant invader known to most of the tropical world. Not only can the quantitative phase information provided by the DHM technique readily relate to chromatin content of the individual cells, which in turn reflects viability. Low viability among invasive cultivars can be used to identify the most suitable horticultural varieties for ornamental use and distribution.  More importantly, from our own research perspective, this work can be expanded to reflect ploidy levels, by means of estimation of C-value content.  

There are several different ways in which this preliminary study can be extrapolated further; at present we have only used the mean phase from among 30 different parameters that can be extracted from the DHM phase images. We are currently in the process of using other parameters for estimating C value content within the grain, through deep learning if possible, which in turn would provide a handle to determine/quantify ploidy levels, an area that could transform plant research, specially for the invasive species where variation in ploidy levels hav e been shown to be an important factor in invasivity. The present work has been conducted at 40X resolution and can be scaled up in principle by expanding the area of examination by lowering the resolution of the brightfield microscope. We are currently exploring the option of using 10X, and widening the area since all the phase features being extracted are already being captured and do not depend on resolution. In summary, the present work on establishing the effectiveness of quantitative phase imaging modality for pollen imaging opens up multiple new avenues for further work as we hope to report in future. 

\section*{Acknowledgements}
We acknowledge support from Central Research Facility, IIT Delhi for making the Digital Holographic Microscope (DHM) system available for this work. Authors also acknowledge the support of National Institute of Plant Genome Research (NIPGR), New Delhi for infrastructure, the Confocal facility of NIPGR for the SEM work and DBT-eLibrary Consortium (DeLCON) for providing access to e-resources. A version of this article is available online as a preprint in arXiv (Kumar et al., 2022).

\section*{Author Contributions}
Conceptualization, KK \& GY; Data Collection: VK, NG, AP; Visualization, KK, VG, NG, GY; Investigation, VK, NG, KK, GY; Writing- Original draft, KK, GY; Writing- Review \& Editing- KK, AP, SB, GY; Funding, GY, KK, AP; Supervision, KK, SB, GY. All authors have read and approved the manuscript.

\section*{Financial Support}
GY acknowledges support of National Institute of Plant Genome Research (NIPGR), New Delhi, and funds from BBSRC GCRF Grant ID. BBSRC BB/P027970/1TIGR2ESS for this work. VK was funded by the same BBSRC grant as a Post Graduate Research Fellow (PGRF). NG acknowledges support from the Prime Minister's Research Fellowship (PMRF) for pursuing PhD thesis work at IIT Delhi. AP received JRF And SRF fellowship from Council of Scientific \& Industrial Research (CSIR), Government of India as part of his PhD at NIPGR. KK acknowledges support from Abdul Kalam Technology Innovation National Fellowship provided by Indian National Academy of Engineering (INAE). The publication charge of this article is covered by NIPGR Core Grant. These funding bodies provided the equipment required for this work but do not have any role in design of the study and collection, analysis, and interpretation of data and in writing the manuscript. 

\section*{Conflicts of Interest Declaration}
Conflicts of interest: None

\section*{Data and Coding Availability Statement}
Bright-field and phase image data used in this work may be made available to users upon reasonable request.

\bibliographystyle{apacite}  
\bibliography{references}  







\end{document}